\documentclass[
reprint,
superscriptaddress,
amsmath,amssymb,
aps,
prl,longbibliography
]{revtex4-1}
\usepackage{float}
\usepackage{xcolor}
\usepackage{graphicx}
\usepackage{color}
\usepackage{amssymb}
\usepackage{amsfonts}
\usepackage{amsmath}
\usepackage{bm}
\usepackage{bbm}

\begin{document}

\title{The Berezinskii-Kosterlitz-Thouless Transition and\\ Anomalous Metallic Phase in a Hybrid Josephson Junction Array}

\author{C.~G.~L.~B\o{}ttcher}\thanks{Present Address: Department of Applied Physics, Yale University, New Haven, CT, 06520, USA. \\
Email: charlotte.boettcher@yale.edu}
\affiliation{Center for Quantum Devices, Niels Bohr Institute, University of Copenhagen, 2100 Copenhagen, Denmark}
\author{F.~Nichele}\thanks{Present Address: IBM Research Laboratory, Z\"urich, Z\"urich, Switzerland}
\affiliation{Center for Quantum Devices, Niels Bohr Institute, University of Copenhagen, 2100 Copenhagen, Denmark}
\author{J.~Shabani}\thanks{Present Address: New York University, New York, NY 10003, USA}
\affiliation{California NanoSystems Institute, University of California, Santa Barbara, CA 93106, USA}
\author{C.~J.~Palmstr\o{}m}
\affiliation{California NanoSystems Institute, University of California, Santa Barbara, CA 93106, USA}
\affiliation{Department of Electrical Engineering, University of California, Santa Barbara, CA 93106, USA}
\affiliation{Materials Department, University of California, Santa Barbara, CA 93106, USA}
\author{C.~M.~Marcus}
\affiliation{Center for Quantum Devices, Niels Bohr Institute, University of Copenhagen, 2100 Copenhagen, Denmark}

\date{\today}
\begin{abstract}
{\footnotesize We investigate the Berezinskii-Kosterlitz-Thouless (BKT) transition in a semiconductor-superconductor two-dimensional Josephson junction array. Tuned by an electrostatic top gate, the system exhibits separate superconducting (S), anomalous metal (M*), and insulating (I) phases, bordered by separatrices of the temperature-dependent of sheet resistance, $R_{s}$. We find that the gate-dependent BKT transition temperature falls to zero at the S-M* boundary, suggesting incomplete vortex-antivortex pairing in the M* phase. In the S phase, $R_{s}$ is roughly proportional to perpendicular magnetic field at the BKT transition, as expected, while in the M* phase $R_{s}$ deviates from its zero-field value as a power-law in field with exponent close to 1/2 at low temperature. An in-plane magnetic field eliminates the M* phase, leaving a small scaling exponent at the S-I boundary, which we interpret as a remnant of the incipient M* phase.}
\end{abstract}
\maketitle 
\indent

Josephson junction arrays (JJAs) have for decades provided model systems for investigating classical and quantum phase transitions with competing ground states, frustration, and complex dynamics \cite{Newrock.1999, Fazio.2001}, 2D superconductivity \cite{Gantmakher.2010, Lin.2015}, and more recently as a basis for quantum simulatation \cite{King.2018}, quantum matter \cite{Leykam.2018}, and protected quantum information \cite{Ioffe.2002, Doucot.2012}. 
It is generally accepted that JJAs exhibit a quantum phase transition between superconducting and insulating phases controlled by the ratio $E_{C}/E_{J}$ of charging energy, $E_{C}$, of a single island to the Josephson energy, $E_{J}$, between neighboring islands. At the superconductor-insulator transition (SIT), the 2D sheet resistance $R_{s}$ of the JJA is roughly the resistance quantum, $R_{s}\sim R_{Q}  \equiv h/4e^{2}$ \cite{Chakravarty.1987, Fisher.1990}, or equivalently, $E_{C}/E_{J} \sim 1$, with a temperature-independent separatrix at $R_{s}\sim R_{Q}$ \cite{Newrock.1999, Fazio.2001}. 

However, this conventional SIT picture misses a commonly observed regime seen in a variety of materials---the anomalous metal---where $R_s$ saturates at low temperature at a tunable value $R_{s} < R_{Q}$ \cite{Kapitulnik.2019}
The origin and requirements  for the anomalous metal are not known, despite years of investigation and speculation \cite{Kapitulnik.2001, Phillips.2003, Kapitulnik.2019,Diamantini.2020, Sacepe.2020}.

\begin{figure}[t]
	\includegraphics[width= 2.9 in]{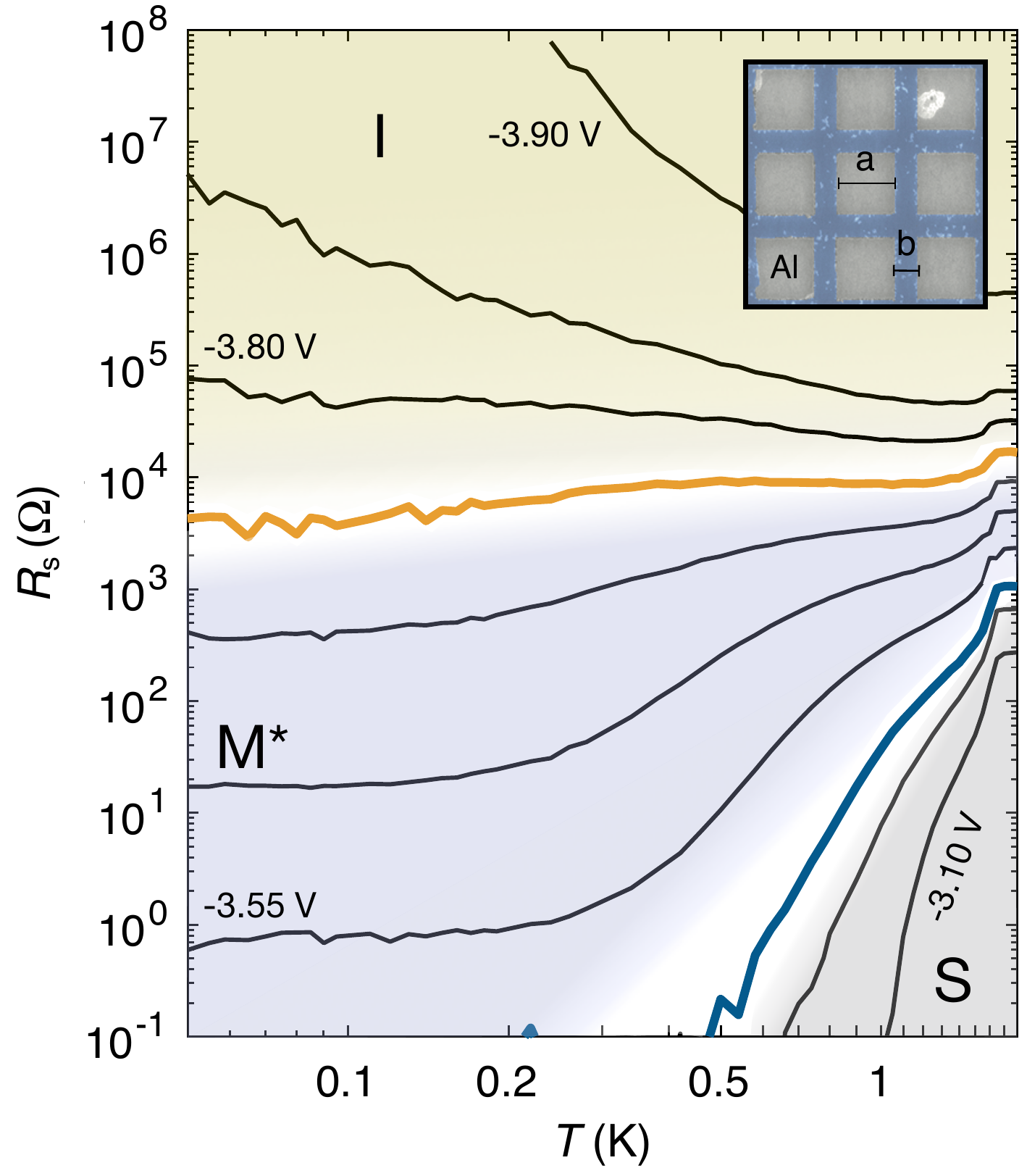}
	\caption{Sheet resistance, $R_{s}(T)$, of the InAs/Al Josephson junction array (inset) as a function of temperature, $T$, at several gate voltages, $V_{G}$, ranging from $-3.1$~V to $-3.9$~V, from Ref.~\cite{Boettcher.2018}. Two separatrices (blue and orange curves) mark the boundaries between superconducting (S) phase, where $R_{s}(T)$ is concave down, becoming unmeasurably small at low $T$, the anomalous metal (M*), where $R_{s}(T)$ is concave up and saturating at low $T$, and the insulating phase, where $R_{s}(T)$ increases with falling $T$. The S-M* boundary (blue) starts at normal-state sheet resistance, $R_{N}\sim 1$~k$\Omega$. The M*-I (orange) is roughly temperature independent, with $R_{s} \sim 6$~k$\Omega$, similar to with a conventional S-I boundary. Inset: false-color electron micrograph with scale bar showing the array before the deposition of the top insulator and metallic gate, with 1~$\mu$m Al squares patterned onto an InAs substrate, as described in Ref.~\cite{Boettcher.2018}. 		}
	\label{fig1}
\end{figure}

At zero magnetic field, temperature destroys 2D superconductivity through a Berezinskii-Kosterlitz-Thouless (BKT) transition, characterized by the unbinding of vortex-antivortex pairs when the temperature, $T$, exceeds a critical value, $T_{\rm BKT}$ \cite{Jose.2013, Kosterlitz.2016, Drouin.2022}. Dissipation by the motion of unbound vortices above $T_{\rm BKT}$ results in nonzero resistivity \cite{Halperin.1979, Kadin.1983}. 

In this Letter, we experimentally investigate the interaction of the BKT transition with the anomalous metal in an InAs/Al heterostructure patterned into a regular array of micron-size Al islands separated by narrow stripes where the Al has been removed, which can be depleted by a global gate. We identify two distinct boundaries where $R_{s}(T)$ curves separate, controlled by gate voltage, defining three regions of superconducting (S), anomalous metal (M*) and insulating (I) phases. 
A similar conclusion was recently proposed for a field-driven SIT \cite{Zhang.2021, Zhang.2022}, and contrasts interpretations where BKT crosses to a quantum-dominated regime throughout the superconducting phase \cite{Lin.2012}, or where inadequate cooling is responsible for anomalous metal behavior \cite{Park.2017, Tamir.2019}.

We find that in S phase, that is, the low-resistance side of the S-M* separatrix, the temperature dependent sheet resistance, $R_{s}(T)$, is well described by the BKT form \cite{Halperin.1979} over three orders of magnitude of $R_{s}$, yielding a gate-voltage--dependent $T_{\rm BKT}$ as a fit parameter.
Importantly, we find $T_{\rm BKT}$ goes to zero at the S-M* boundary, {\it not} at the $T$-independent separatrix (our M*-I boundary) where $R_{s} \sim R_{Q}$, as one would expect for a conventional SIT \cite{Fisher.1986, Fazio.2001}.  
The observation of a vanishing $T_{\rm BKT}$ at the S-M* boundary suggests that the transition from M* to the normal metal (M) with increasing temperature is not BKT-like. We interpret the result as incomplete vortex-antivortex pairing that persists to zero temperature in the M* phase. 

We next demonstrate power-law dependences of $R_{s}$ on small perpendicular field, $B_{\perp}$, in both the S and M* phases. Linear magnetoresistance, $R_{s}(B_{\perp},T_{\rm BKT})\propto B_{\perp}^{\beta}$, with $\beta = 1$, is expected around the BKT transition \cite{Chen.1995, Chen.1996, Mason.2001, Saito.2015, Chen.2021}. We find that in the S phase, $\beta(T) \sim 1$ near $T_{\rm BKT}$, though we find $\beta(T)$ increasing with $T$, contrary to \cite{Chen.1995, Chen.1996}. In the M* phase, beyond where $T_{\rm BKT}\rightarrow 0$, we again find a power law after subtracting the saturation value, $[R_{s}(B_{\perp},T)-R_{s}(0,T)]\propto B_{\perp}^{\beta}$, though with $\beta < 1$, approaching $\beta = 1/2$ well into M* [see Figs.~\ref{fig3}(b,c)]. 

Finally, we find that a moderate in-plane magnetic field, $B_{\parallel} \sim 0.5$~T, known to eliminate M* \cite{Boettcher.2018}, causes the S-M* and M*-I phase boundaries to coalesce into to a single broad S-I boundary (Fig.~\ref{fig4}). The broad S-I transition at $B_{\parallel}=0.5$~T shows a small scaling exponent at low-$T$, a possible remnant of the M* phase. 
A transition from slow and fast temperature dependence is seen toward the S side of the broad S-I transition, separated from the S-I crossing point at $R_{s}\sim R_{Q}$, presumably another vestige of the M* phase.  
The JJA was fabricated using a hybrid InAs/Al heterostructure, with a 7~nm InAs quantum well separated from a 7~nm epitaxial Al surface layer by 10~nm InGaAs barrier \cite{Boettcher.2018}. A Hall bar was patterned by wet etching through the quantum well. A subsequent patterning of 40x100 array of $1 \mu$m Al squares separated by 150~nm (Device A) or 350 nm (Device B) spacing was patterned by removing the Al between squares using Transene D aluminum etchant (see Fig.~\ref{fig1}, inset). The array was then covered by a 40 nm Al$_{2}$O$_{3}$ insulating layer followed by a Ti/Au top gate. The two measured devices behaved similarly. Except where noted data are from Device A.

Devices were measured in a dilution refrigerator with a base mixing chamber temperature of 20 mK. 
A vector magnet was used to independently apply perpendicular and in-plane magnetic fields applied along the current direction, after calibrating the  magnet axes to compensate for small sample tilt. 
A four-wire measurement of longitudinal resistance with both current and voltage measured was carried out using standard AC lock-in techniques, keeping the voltage across the array below $5 \mu$V. 2D sheet resistance $R_{s}(T)$, spanning $0.1\,\Omega$ to $100\,{\rm M}\Omega$, were accessible by tuning the top-gate voltage, $V_{G}$, in the range $-3$~V to $-4$~V for both devices.  Over this same range, normal-state sheet resistance $R_{N}$, measured above the critical temperature  $T_{c0}\sim 1.6$~K of the Al islands, spanned $R_{N}\sim 100\,\Omega$ at  $V_{G}\sim -3$~V to $R_{N}\sim 1\, {\rm M}\Omega$ at $V_{G}\sim -4$~V, as shown in Fig.~\ref{fig1}.

The experimental phase diagram in the $T$-$V_{G}$ plane is shown in Fig.~\ref{fig1}, along with representative $R_{s}(T)$ curves at fixed $V_{G}$. In the superconducting (S) region, with $R_{N}\lesssim 1\, {\rm k}\Omega$, all $R_{s}(T)$ curves were found to decrease with lower $T$ down to the smallest measurable resistance, $R_{s}\sim 0.1\,\Omega$. At more negative gate voltages, in the region marked M*, $R_{s}$ initially falls, then saturates at a $V_{G}$-dependent value ranging from $\sim 1\,\Omega$ up to $\sim R_{Q}$. As discussed below, $R_{s}(T)$ curves in M* are distinct from those in S throughout the temperature range. In particular, $R_{s}(T)$ in M* is not well described by the BKT form \cite{Halperin.1979}. Comparable to the conventional SIT, the M*-I separatrix (orange curve in Fig.~\ref{fig1}) occurs at $R_{s}\sim R_{Q}$ and is roughly independent of $T$ from lowest measured temperature up to $T_{c0}$.  At more negative $V_{G}$ beyond the M*-I separatrix in the region marked I, $R_{s}(T)$ rises with lower $T$, characteristic of the insulating phase. We have previously investigated variable range hopping and activated regimes of the I phase, and did not observe low-$T$ saturation I phase away from M*-I separatrix \cite{Boettcher.2018}. Note in Fig.~\ref{fig1} that $R_{N}$ slightly exceeds $R_{s}(T)$ just below $T_{c0}$ throughout the measured S, M*, and I phases, so that the transition at $T_{c0}$ makes a small upward step in resistance going from below to above the transitions where the islands become normal. The critical temperature for the islands,  $T_{c0}\sim 1.6$~K, is independent of $V_{G}$.

Representative $R_{s}(T)$ curves are shown in Fig.~\ref{fig1}. For a larger set of curves, finely sampled in $V_{G}$ throughout the S and M* phases, we fit $R_{s}(T)$ to the BKT form \cite{Halperin.1979, Kadin.1983},  

\begin{equation}
R_s(T) = a R_{N}\exp\Big[-b(\frac{T_{c0}-T}{T-T_{\rm BKT}})^{1/2}\Big].
\label{HN}
\end{equation}

\begin{figure}
	\includegraphics[width= 3.2 in]{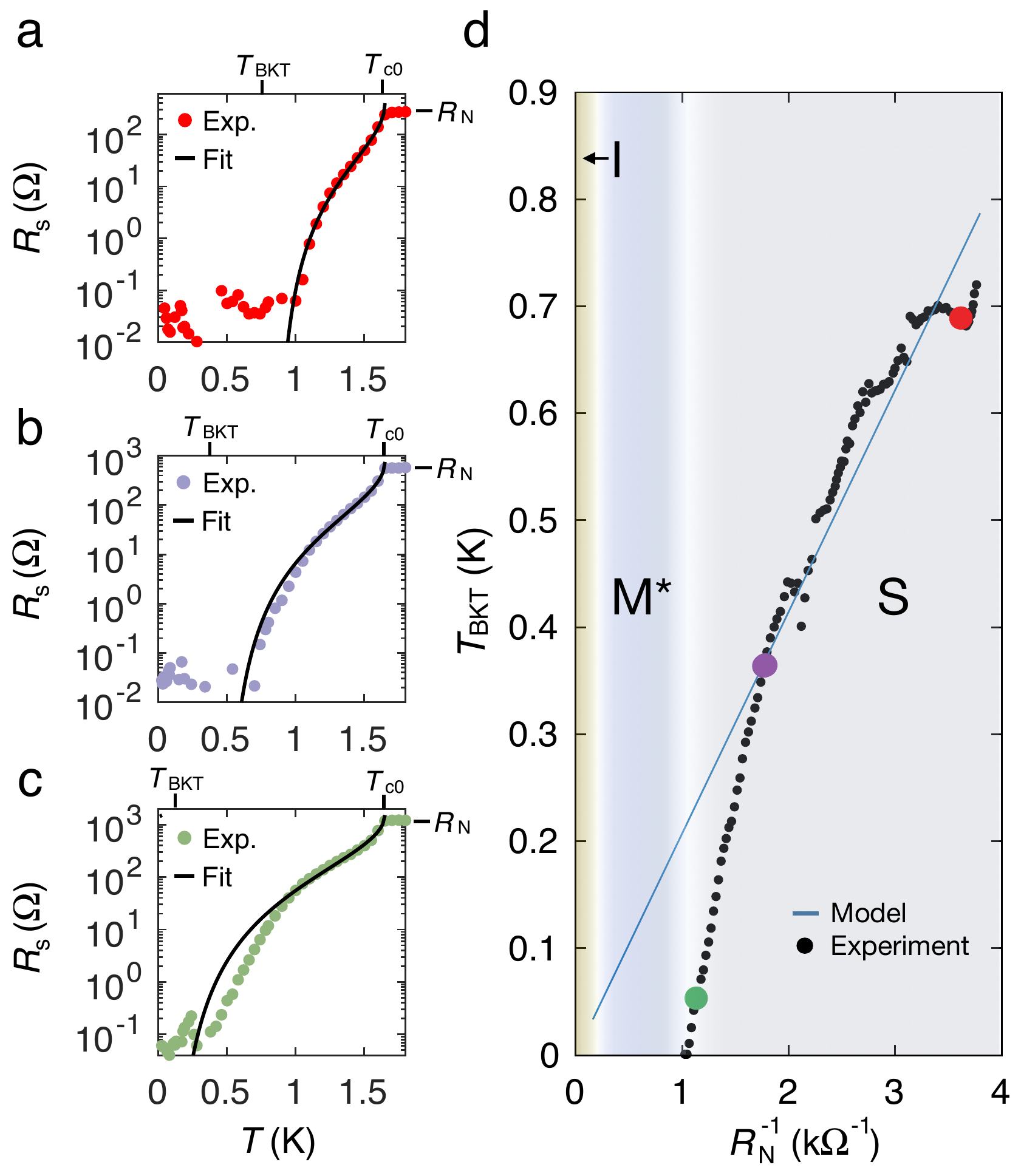}
	\caption{(a-c) Sheet resistance, $R_{s}(T)$ as a function of temperature, $T$ at three gate voltages (filled colored circles), parameterized by normal-state sheet resistance, $R_{N}$, along with fits to Eq.~\ref{HN}. Top axes show gate-independent critical temperature, $T_{c0}$ of the Al squares, read directly from the data, and BKT transition temperature, $T_{\rm BKT}$ from the fits. Note the good fit deep in S, which becomes poorer toward the S-M* boundary. (d) BKT transition temperature $T_{\rm BKT}$ extracted from fits at many gate voltages (black filled circles), as a function of $R_{N}^{-1}$. $T_{\rm BKT}$ reaches zero near the S-M* boundary, defined by the blue separatrix in Fig.~\ref{fig1}. Three colored data points correspond to panels a-c. Model line is based on linear relations $T_{\rm BKT} \sim E_{J} \propto I_{c} \propto R_{N}^{-1}$, with a single fit parameter (see text). }	
	\label{fig2}
\end{figure}

\begin{figure}
	\includegraphics[width= 3 in]{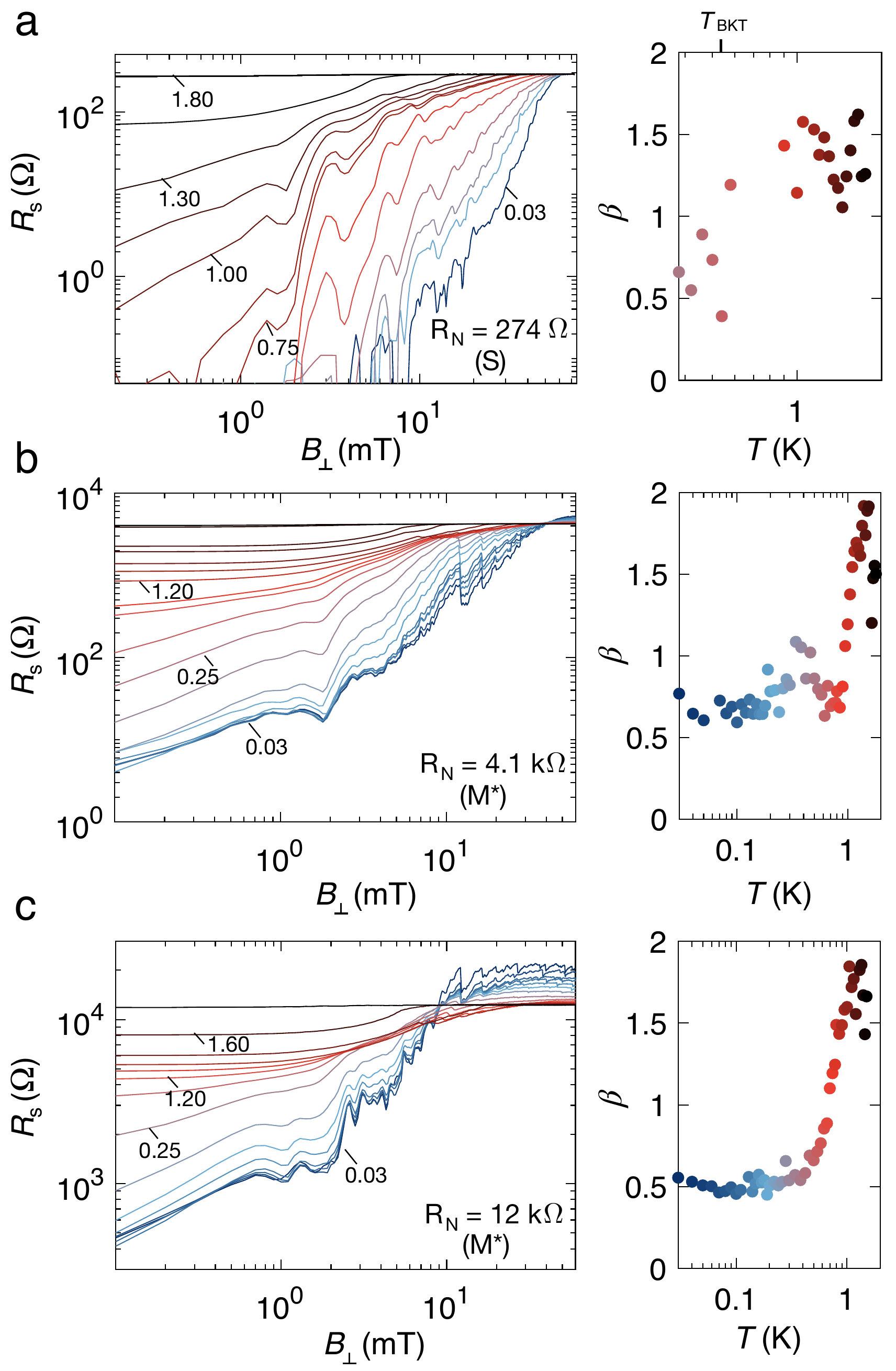}
	\caption{Low-field magnetoresistance $R_{s}(B_{\perp}, T)$ as a function of perpendicular magnetic field, $B_{\perp}$, shows a power-law dependence (straight line on a log-log plot), $R_{s}(B_{\perp}, T)- R_{s}(0, T) = A(T) B_{\perp}^{\beta(T)}$  with a power $\beta(T)$ dependence on temperature, $T$ (marked on graphs), and gate-controlled normal-state sheet resistance, $R_{N}$. (a) In the superconducting (S) phase, on the low-resistance (less negative $V_G$) side of the S-M* boundary, with $R_{N}=274\, \Omega$, $\beta(T_{\rm BKT}) \sim 1$. (b-c) In M*, $\beta(T)< 1$ at low $T$, and appears to settle around $\beta \sim 1/2$. Fluctuations in $R_{s}(B_{\perp})$ on a $\sim 1$~mT scale result from flux commensuration effects.}
	\label{fig3}
\end{figure}

\begin{figure}[t]
	\includegraphics[width= 2.65 in]{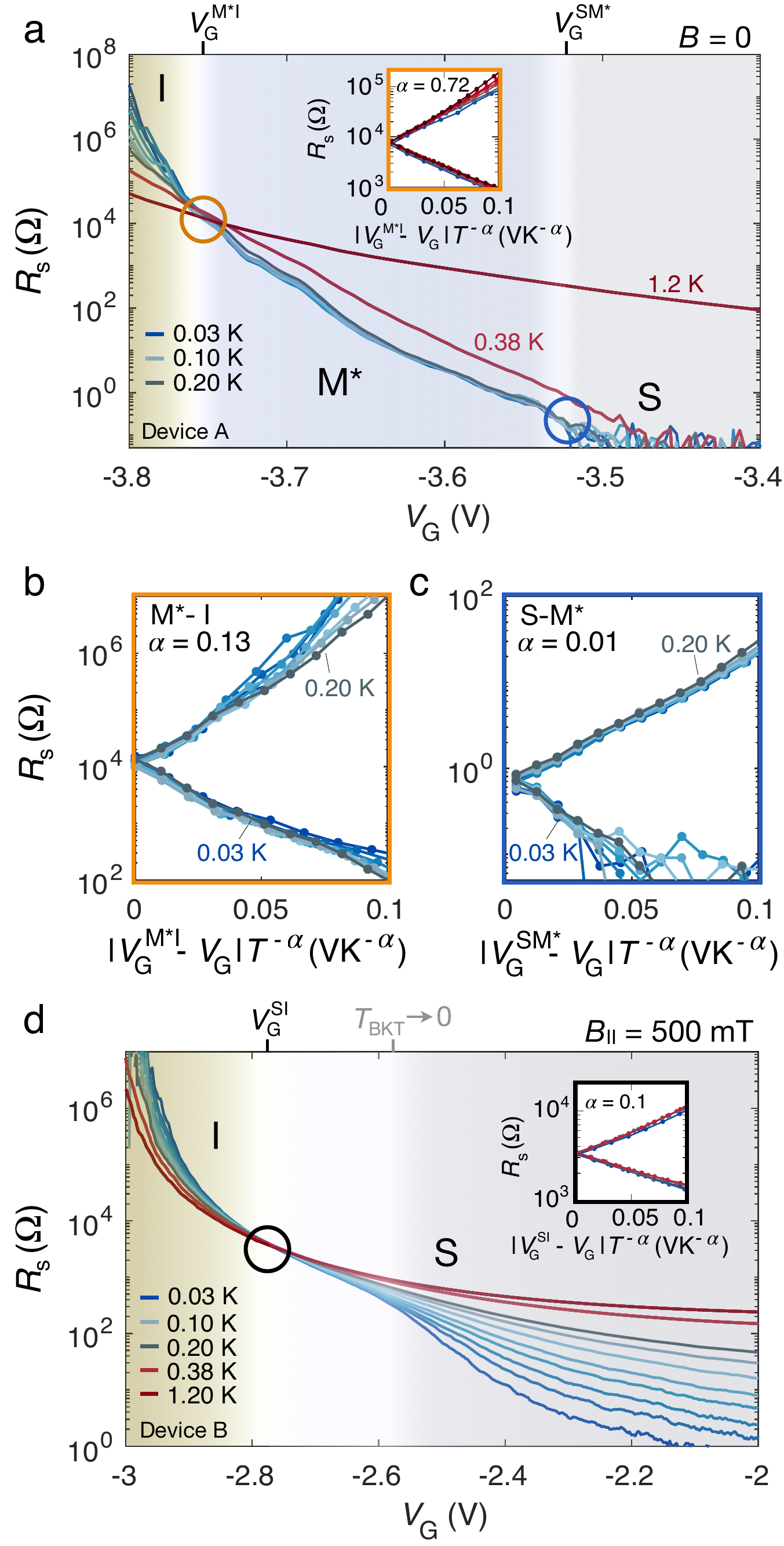}
	\caption{(a) Sheet resistance, $R_{s}$, as a function of gate voltage, $V_{G}$ at $B_{\parallel} = B_{\perp} =0$ shows $T$-independent curves at low $T$ across the anomalous metal phase, M*, which flair out for different $T$ in the superconducting (S) and insulating (I) phases.  Inset: Scaling using only high-$T$ isotherms ($T>0.2$~K) at the M*-I boundary (orange circle in main panel) yields  exponent $\alpha = 0.72$. (b) Scaling using only low-$T$ isotherms ($T<0.2$~K) at the M*-I boundary yields $\alpha = 0.13$. (c) Scaling at the S-M* boundary [blue circle in (a)], yields $\alpha = 0.01$. (d) Applying an in-plane magnetic field, $B_{\parallel} = 0.5$~T, results in a single crossing of isotherms (black circle). Inset: Scaling with all isotherms yields $\alpha = 0.1$. We interpret the small $\alpha$, i.e., nearly parallel isotherms, as a vestige of M*, where low-$T$ isotherms coincide. The broad S-I transition (white region) is bounded by the crossing point on the I side and the rapid spreading of isotherms on the S side. The top axis indicates where $T_{\rm BKT} \rightarrow 0$ following similar analysis to Fig.~\ref{fig2}. Data for Device B.
			}
	\label{fig4}
\end{figure}

For each $V_{G}$, a least-square fit to the logarithm of $R_{s}(T)$ yields $a$, $b$, and $T_{\rm BKT}$ as fit parameters, with $T_{c0} = 1.6$~K and $R_{N}(V_{G})$ taken from measurements. Figure \ref{fig2}(a) shows a typical fit deep in the S regime, yielding excellent agreement between experimental and Eq.~\ref{HN} over several orders of magnitude of $R_{s}(T)$ from the lowest measured resistance to $R_{N}$. The resulting $T_{\rm BKT}$ is indicated along the top axis. As $V_{G}$ is set more negative but still within S, the data and fits begin to deviate most noticeably for $T\lesssim 1$~K. Note that the measurement lies {\it below} the fit, in contrast to \cite{Lin.2012}. The extracted $T_{\rm BKT}$ values, marked on the top axes of Figs.~\ref{fig2}(b,c), move quickly toward zero as the S-M* boundary is approached at more negative gate voltages.

Repeating fits across a range of gate voltages within the S region yields the values for $T_{\rm BKT}$ shown in Fig.~\ref{fig2}(d), plotted as a function of $R_{N}^{-1}$, along with a classical model (absent charging effects), $T_{\rm BKT}\propto R_{N}^{-1}$. 
We find that $T_{\rm BKT}$ is roughly proportional to $R_{N}^{-1}$ in the S region, but deviates below the model line, reaching $T_{\rm BKT}=0$ at a value of $R_{N}(V_{G})\sim 1\,{\rm k}\Omega$, coinciding with the S-M* boundary. 
We emphasize that value of $V_G$ marked as the the S-M* boundary in Fig.~\ref{fig2} was defined by the separatrix (blue trace) in Fig.~\ref{fig1}, not by the point where $T_{\rm BKT}=0$. 
The observation that $T_{\rm BKT}$ reaches zero at the S-M* boundary is a striking experimental result, not expected within a conventional SIT picture, where $T_{\rm BKT}\rightarrow0$ at $R_{s} \sim R_{Q}$, i.e., the M*-I boundary in this case.

The simple proportionality $T_{\rm BKT}=\gamma R_{N}^{-1}$ follows from the Ambegaokar-Baratoff relation at $T \ll \Delta$ for individual junctions, $I^{(i)}_{c}R^{(i)}_{N}=\pi\Delta/2e$, where $\Delta$ is the superconducting gap induced in the InAs under the Al islands. Setting $R_{N}=R^{(i)}_{N}$ and $E_{J} = (\hbar/2e)I^{(i)}_{c}$  gives $E_{J}=(\Delta/2)(R_{Q}/R_{N})$. 
The classical BKT relation, $k_{B}T_{\rm BKT}=(\pi/2)E_{J}$ \cite{Tinkham.2004} can then be written $T_{\rm BKT} = (\pi\Delta/4k_{B})(R_{Q}/R_N)$. 

An upper bound on $\Delta$ based on the critical temperature of the Al islands gives $\Delta \lesssim 1.76\, k_{B} T_{c0} = 240\,\mu$eV, using the measured $T_{c0}=1.6$~K. A lower bound based on the $I_cR_N$ product, using the array depinning current $I_d^{(A)} = 20\,\mu {\rm A} \lesssim 40\,I_c^{(i)}$ of the 40-junction-wide array  at $R_N = 0.4\, {\rm k}\Omega$, gives $\Delta \gtrsim 120\, \mu$eV \cite{Boettcher.2018}. Tunneling spectroscopy into the InAs adjacent to the Al edge in similar material yields $\Delta \sim 190\,\mu$eV \cite{Kjaergaard.2016}. Using this value gives $\gamma = (\pi\Delta/4k_{B})R_{Q} \sim 11\,{\rm k}\Omega\,$K.

Experimentally, we observe the proportionality $T_{\rm BKT}=\gamma^{(\rm{exp})} R_{N}^{-1}$ in the S phase, as seen in Fig.~\ref{fig2}(d), but with a considerably smaller slope, $\gamma^{\rm (exp)} \sim 0.2\,{\rm k}\Omega\,{\rm K}$. We interpret the discrepancy as reflecting a suppressed BKT transition temperature, $T_{\rm BKT} \sim 0.02\, E_{J}$---while keeping $E_{J}\sim(\Delta/2)(R_{Q}/R_{N})$---instead of the classical relation $T_{\rm BKT}\sim E_J$. 
This interpretation is consistent with a charging energy $E_{C} \sim 150\, \mu e {\rm V}$, previously measured in the I phase \cite{Boettcher.2018}, being comparable to $E_J$ when $R_{s}\sim R_N\sim R_{Q}$. 


As $V_{G}$ becomes more negative, moving the system from the S phase toward the S-M* boundary, $T_{\rm BKT}$ deviates from the proportionality $T_{\rm BKT}=\gamma R_{N}^{-1}$, reaching zero (roughly linearly) at the S-M* boundary, as seen in Fig.~\ref{fig2}(d). In conventional SIT systems, reduction of $T_{\rm BKT}$ below $\sim E_{J} \propto R_{N}^{-1}$ is an expected and well investigated consequence of quantum corrections associated with island capacitance \cite{Jose.1994, Zant.1996, Newrock.1999, Fazio.2001}. However, in the conventional case, the suppressed $T_{\rm TBK}$ reaches zero at the SIT, where $R_{N}\sim R_{Q}$. In the present case, one might have expected a deviation of $T_{\rm BKT}$ to reach zero at the corresponding M*-I separatrix where $R_{s} \sim R_{N} \sim R_{Q}$, but that is not what is observed. Instead, we find that $T_{\rm BKT}$ reaches zero at the S-M* boundary, and that $T_{\rm BKT} = 0$ throughout the M* phase.

These observation, the 50-fold reduction in slope of $T_{\rm BKT}$ versus $R_{N}^{-1}$ in the S phase and the vanishing of $T_{\rm BKT}$ in the M* phase, suggest a picture of vortex-antivortex binding that is suppressed in S and fails in M*. As discussed below, this picture is consistent with the observed scaling in M* of $R_s$ with small perpendicular magnetic field.

The observed vanishing of $T_{\rm BKT}$ at the S-M* boundary is consistent with another method of determining $T_{\rm BKT}$: a jump in the voltage-current characteristic, $V \propto I^{\zeta}$, from ohmic, $\zeta = 1$, for $T>T_{\rm BKT}$ to $\zeta = 3$ at $T=T_{\rm BKT}$. The condition $\zeta = 3$ is equivalent to the point where $R_{s}\,\,(\propto dV/dI\propto V^{\zeta-1})$ first touches zero at its minimum at zero bias. Because $R_{s}$ is an even function of $I$, when it first touches zero it will in general be parabolic to lowest order, $R_{s}\propto I^{2}$, which is equivalent to $V\propto I^{3}$. Throughout M*, the nonzero $R_{s}$ down to the lowest temperatures implies ohmic response, $\zeta=1$. The condition $\zeta = 3$ is never reached in M*, consistent with $T_{\rm BKT}=0$. 

By the same argument, now in the S phase, $T_{\rm BKT}$ is roughly where $R_{s}\,\,(\propto V^{\zeta-1})$ first touches zero. That is, $R_s$ will be parabolic ($\zeta-1 \sim 2$) at $T_{\rm BKT}$. Below $T_{\rm BKT}$, $dV/dI$ will be a flat-bottomed, even function around zero bias, giving $\zeta \geq 5$. Above $T_{\rm BKT}$, the ohmic $R_{s}$ gives $\zeta= 1$. This implies a jump from $\zeta = 1$ to $\zeta \sim 3$ at $T_{\rm BKT}$ in the S phase.

We next examine magnetoresistance $R_{s}(B_{\perp})$ at small perpendicular magnetic field, $B_{\perp}$, which has been investigated previously to identify the BKT transition in arrays and 2D films. 
Conventional SITs ~\cite{Martin.1989, Chen.1996, Sambandamurthy.2006} and anomalous metals~\cite{Tsen.2016, Chen.2021, Zhang.2021, Zhang.2022} often show a power-law dependence, $R_{s}(B_{\perp}, T)- R_{s}(0, T) = A(T) B_{\perp}^{\beta(T)}$, in some cases consistent the BKT prediction, $\beta(T_{\rm BKT}) = 1$ \cite{Martin.1989, Chen.1996, Chen.2021}. 
Other experiments \cite{Sambandamurthy.2006} find $\beta(T) >1$, consistent with activated vortex creep with logarithmic vortex interactions \cite{Feigelman.1990}. 
In systems exhibiting anomalous metal behavior [$R_{s}(0, 0)\ne 0$], magnetic field dependence of the low-$T$ saturating resistance was observed to be activated \cite{Ephron.1995} or power-law with $1 < \beta(T) < 3$ in MoGe films \cite{Wu.2006} and crystalline NbS$_{2}$ \cite{Tsen.2016}. 
In higher-resistance granular InO$_{2}$ films, smaller exponents were found, $\beta(T\rightarrow0)) = 0.66$ \cite{Zhang.2021} and $\beta(T\rightarrow 0) = 0.45$ \cite{Zhang.2022} in granular films. Reference~\cite{Li.2019} reports $\beta(T) \sim 2$, also consistent with \cite{Wu.2006}, though their lowest temperature data appears to show $\beta(T) < 1$. 

Figure \ref{fig3} shows $R_{s}(B_{\perp}, T)$ at several temperatures,  in the S and M* phases, along with best-fit values for $\beta(T)$ for each $T$ to the form $R_{s}(B_{\perp}, T)- R_{s}(0, T) = A(T) B_{\perp}^{\beta(T)}$, with prefactor $A(T)$. The power-law can be seen at the lowest fields, less than one flux quantum per plaquette [$(h/2e)({\rm a+b})^{-2} \sim 2$~mT, see Fig.~\ref{fig1} inset], modulated by flux-commensuration effects above $\sim 2$~mT. In the S phase [Fig.~\ref{fig3}(a)], $\beta(T) \sim 1$ for $T \sim T_{\rm BKT}\sim 0.5$~K, roughly consistent with results in Fig.~\ref{fig2}(d) for similar $R_N$. 
We note without explanation that the trend of increasing $\beta(T)$ with temperature is opposite of the trend reported in \cite{Chen.1996}. In the S phase, $R_{s}(B_{\perp})$ for $T < T_{\rm BKT}$ is below the experimental noise floor. 

Throughout the M* phase, where $T_{\rm BKT} = 0$, we find $\beta(T) < 1$ at low-temperatures (noting a bump to $\beta \sim 1$ around 0.5~K in Fig.~\ref{fig3}(b), close the S-M* boundary). Near the M*-I border, $\beta\sim 1/2$ for a broad range of temperatures similar to \cite{Li.2019, Zhang.2022}. 

We speculate that the observed $\beta$ reflects incomplete binding of vortex-antivortex pairs in M*. Within this picture, the value of coupling constant $K=J/k_BT$ where vortex-antivortex pairs form, but fail to bind, is $1/\pi$ ~\cite{Solla.1981, Kosterlitz.2016}. Then, the relation $\beta=\pi K/2$ \cite{Chen.1996} gives $\beta=1/2$. In the S phase, on the other hand, vortex-antivortex pairs bind at $T_{\rm BKT}$, giving $\beta=1$.

Applying an in-plane magnetic field, $B_{\parallel}$ suppresses the M* phase, restoring a conventional SIT, though with a small and field-dependent scaling exponent, as shown in Fig.~\ref{fig4}. In Ref.~\cite{Boettcher.2018}, we speculated that the suppression of the M* phase with in-plane field results from increased dissipation from the soft superconducting gap, which stabilizes phase fluctuations, in this case caused by the softening of the induced gap by the in-plane field \cite{Kjaergaard.2016, Banerjee.2019}. This interpretation differs from Ref.~\cite{Kapitulnik.2001}, which found that galvanic (though not capacitive \cite{Mason.2002}) coupling to a dissipative channel induces rather than suppressed the anomalous metal phase. Stabilization of superconductivity with in-plane field has been reported in a variety of 2D systems, including LaAlO$_{3}$/SrTiO$_{3}$ interfaces \cite{Gardner.2011} as well as Pb \cite{Gardner.2011, Niwata.2017} and WTe$_{2}$ \cite{Asaba.2018} thin films. Stabilization of superconductivity by an in-plane field has been attributed to mechanisms besides dissipation, including the compensation \cite{Jaccarino.1962} and freezing \cite{Kharitonov.2005} of magnetic impurities. Similar stabilization of superconductivity by an applied magnetic field was also found in 1D nanowires, where it was also attributed to dissipation induced by the applied field \cite{Tian.2005, Fu.2006}. 
Scaling analysis for conventional SITs can be applied to a gate-voltage controlled transition using the scaled voltage axis, $|V_{G}^{S\rm{-}I} - V_{G}|T^{-\alpha}$, with scaling exponent $\alpha = (z\nu)^{-1}$ where $\nu$ and $z$ are spatial and dynamical exponents \cite{Fazio.2001, Boettcher.2018}. Using $B_{\parallel}$ to eliminate the M* phase, we examine $\alpha(B_{\parallel})$ of the S-M*, M*-I (Fig.~\ref{fig4}(a-c)) and S-I (Fig.~\ref{fig4}(d)) transitions. For the M*-I transition (orange circle) using only higher temperature data ($T > 0.2$~K)  yields $\alpha = 0.72$ consistent with classical percolation, $z\nu=4/3$ \cite{Boettcher.2018} while low-$T$ scaling (Fig.~\ref{fig4}(b)), yields $\alpha=0.13$. Scaling at the S-M* boundary, (blue circle), defined by the S-M* separatrix in Fig.~\ref{fig1}, coinciding with the value of $V_G$ where $T_{\rm BKT}\rightarrow 0$, also yields a small value, $\alpha(B_{\parallel}=0) = 0.01$ (Fig.~\ref{fig4}(c)).
The reduced value of $\alpha$ is not surprising since within the M* phase, $dR_{s}/dT\rightarrow 0$ which yields $\alpha \rightarrow 0$ on the M* side of the transition. 

Applying an in-plane field causes the S-M* and M*-I boundaries to coalesce, eliminating a clear M* phase (Fig.~\ref{fig4}(d)). Scaling at the the remaining single crossing of isotherms yields a small exponent $\alpha(B_{\parallel}=0.5 {\rm T}) \sim 0.10$. At the extended S-I crossing at nonzero $B_{\parallel}$ isotherms diverge only weakly. The single crossing in Fig.~\ref{fig4}(d) is the remnant of the M*-I crossing at $B_{\parallel} = 0$  [orange circle in Fig.~\ref{fig4}(a)], where low-$T$ isotherms diverge on the I side but remain parallel in M*, only to diverge again at the remote S-M* boundary (blue circle). 
We do not interpret the small $\alpha$ as a Griffiths transition \cite{Xing.2015, Liu.2019, Han.2020}, though cannot rule it out. Rather, we interpret the small $\alpha$ as a vestige of the M* phase, where $\alpha$ vanishes on the M* sides of the S-M* and M*-I boundaries.\\

We thank A. Kapitulnik, S. Kivelson, and B. Spivak for useful discussion. Research supported by the Danish National Research Foundation, Microsoft, and a research grant (Project 43951) from VILLUM FONDEN.

\bibliography{bibliography}

\end{document}